\documentclass[twocolumn,showpacs,preprintnumbers,amsmath,amssymb]{revtex4}

\usepackage{dcolumn}% Align table columns on decimal point
\usepackage{bm}% bold math

\begin{document}

\title{Validity of semiclassical gravity in the stochastic gravity approach}

\author{E. Verdaguer}
\email{verdague@ffn.ub.es}

\affiliation{Departament de F\'{\i}sica Fonamental and CER en
Astrof\'\i sica, F\'\i sica de Part\'\i cules i Cosmologia,
Universitat de Barcelona, Av.~Diagonal 647, 08028 Barcelona,
Spain}


\begin{abstract}
In semiclassical gravity the back-reaction of the classical
gravitational field interacting with quantum matter
fields is described by the
semiclassical Einstein equations. A criterion for the validity of
semiclassical gravity based on the stability of the solutions of
the semiclassical Einstein equations with respect to quantum
metric perturbations is discussed. The two-point quantum
correlation functions for the metric perturbations can be
described by the Einstein-Langevin equation obtained in the
framework of stochastic gravity. These correlation functions
agree, to leading order in the large $N$ limit, with the quantum
correlation functions of the theory of gravity interacting with
$N$ matter fields.  The Einstein-Langevin equations  exhibit
runaway solutions and methods to deal with these solutions are
discussed. The validity criterion is used to show that flat
spacetime as a solution of semiclassical gravity is stable and,
consequently, a description based on semiclassical gravity is a
valid approximation in that case.

\end{abstract}

\pacs{04.62.+v, 03.65.Sq, 05.40.-a}

\maketitle

%%%%%%%%%%%%%%%%%%%%%%

\section{Introduction}
\label{sec1}

Semiclassical gravity describes the interaction of the
gravitational field as a classical field with quantum matter
fields. For a free quantum field this theory is robust in the
sense that it is consistent and fairly well understood
\cite{birrell82,wald94}. The gravitational field is described by
the semiclassical Einstein equation which has as a source the
expectation value in some quantum state of the matter stress tensor
operator. The semiclassical theory is in some
sense unique as a theory where the gravitational field is
classical. In fact, a classical gravitational field interacts with
other fields through their stress tensors, and the only reasonable
c-number stress tensor that one may construct
\cite{Wal77,Wal78a,Wal78b} with the stress tensor operator of a
quantum field is its expectation value in some quantum state.
However, the scope and limits of the theory are not so well
understood because we still lack a fully well understood quantum
theory of gravity. It is assumed that the semiclassical theory
should break down at Planck scales, which is when simple order of
magnitude estimates suggest that the quantum effects of gravity
cannot be ignored: the gravitational energy of a quantum
fluctuation of energy in a Planck size region, determined by the
Heisenberg uncertainty principle, is of the same order of
magnitude as the energy of the fluctuation itself.

{}From the semiclassical Einstein equations it seems also clear that
the semiclassical theory should break down when the quantum
fluctuations of the stress tensor are large. Ford \cite{ford82}
was among the first to have emphasized the importance of these
quantum fluctuations. It is less clear, however, how to quantify
the size of these fluctuations. Thus, Kuo and Ford \cite{kuo93}
used the variance of the fluctuations of the stress tensor
operator compared to the mean value as a measure of the validity
of semiclassical gravity. As pointed out by Hu and Phillips
\cite{hu00,phillips00} such a criterion should be refined by
considering the back reaction of those fluctuations on the metric.
Ford and collaborators also noticed that the metric fluctuations
associated to the matter fluctuations can be meaningfully
classified as ``active'' \cite{ford97,yu99,yu00} and ``passive''
\cite{ford82,kuo93,ford99,ford03,borgman03}.

A different approach to the validity of semiclassical gravity was
taken by Horowitz \cite{horowitz80,horowitz81} who studied the
stability of a semiclassical solution with respect to linear
metric perturbations. In the case of a free quantum matter field
in its Minkowski vacuum state, flat spacetime is a solution of
semiclassical gravity. The equations describing those metric
perturbations involve higher order derivatives, and Horowitz found
unstable ``runaway'' solutions that grow exponentially with
characteristic timescales comparable to the Planck time; see also
the analysis by Jordan \cite{jordan87a}. Later, Simon
\cite{simon90,simon91}, argued that those unstable solutions lie
beyond the expected domain of validity of the theory and
emphasized that only those solutions which resulted from
truncating perturbative expansions in terms of the square of the
Planck length are physically acceptable \cite{simon90,simon91}.
Further discussion was provided by Flanagan and Wald
\cite{flanagan96}, who advocated the use of an ``order reduction''
prescription first introduced by Parker and Simon \cite{parker93}.
More recently Anderson, Molina-Par\'\i s and
Mottola have taken up the issue of the validity of semiclassical
gravity \cite{anderson03} again. Their starting point is the fact
that the semiclassical Einstein equation will fail to provide a
valid description of the dynamics of the mean spacetime geometry
whenever the higher order radiative corrections to the effective
action, involving loops of gravitons or internal graviton
propagators, become important. Next, they argue qualitatively that such
higher order radiative corrections cannot be neglected if the
metric fluctuations grow without bound. Finally, they propose a
criterion to characterize the growth of the metric fluctuations,
and hence the validity of semiclassical gravity, based on the
stability of the solutions of the linearized semiclassical
equation. Following these approaches the Minkowski
metric is shown to be a
stable solution of semiclassical gravity with respect to small
metric perturbations.

As emphasized in Ref. \cite{anderson03} the above criteria may be
understood as criteria within semiclassical gravity itself. It is
certainly true that stability is a necessary condition for the
validity of a semiclassical solution, but one may also look for
criteria within extensions of semiclassical gravity. In the
absence of a quantum theory of gravity such criteria may be found
in more modest extensions. Thus, Ford \cite{ford82} considered
graviton production in linearized quantum gravity and compared the
results with the production of gravitational waves in
semiclassical gravity. Ashtekar \cite{Ash96} and Beetle
\cite{Bee98} found large quantum gravity effects in
three-dimensional quantum gravity models. In a recent paper
\cite{HuRouVer04a} (see also Ref. \cite{HuRouVer04b}) we advocate
for a criteria within the stochastic gravity approach. Stochastic
semiclassical gravity extends semiclassical gravity by
incorporating the quantum stress tensor fluctuations of the matter
fields; see Refs. \cite{hu03a,hu04a} for reviews.

It turns out that this validity criteria is equivalent to the
validity criteria that one might advocate within the large $N$
expansion, that is the theory describing the interaction of the
gravitational field with $N$ identical matter fields. In the
leading order, namely the limit in which $N$ goes to infinity and
the gravitational constant is appropriately rescaled, the theory
reproduces semiclassical gravity. Thus, a natural extension of
semiclassical gravity is provided by the next to leading order. It
turns out that the symmetrized two-point quantum correlations of
the metric perturbations in the large $N$ expansion are equivalent
to the two-point stochastic metric fluctuations predicted by
stochastic gravity. Our validity criterion can then be
summarized as follows: a solution of semiclassical gravity is
valid when it is stable with respect to quantum metric
perturbations.  This criterion implies to consider the quantum
correlation functions of the metric perturbations.

It is important to emphasize that the above validity criterion
incorporates in a unified and self-consistent way the two main
ingredients of the criteria exposed above. Namely, the criteria
based on the quantum stress tensor fluctuations of
the matter fields, and the criteria based on the stability of
semiclassical solutions against classical metric perturbations. In
the following discussion we will argue that the former is
incorporated through the so called \textit{induced} fluctuations
and the later though the so called \textit{intrinsic}
fluctuations. These correspond to Ford's ``passive'' and
``active'' fluctuations, respectively. We will see that
symmetrized quantum two-point metric fluctuations can always be
decomposed as a sum of induced and intrinsic fluctuations.

The paper is organized as follows. In section \ref{2} we briefly
review the main ingredients of semiclassical gravity. In section
\ref{3} we introduce stochastic gravity as a theory that goes
beyond semiclassical theory by incorporating the fluctuations of
the quantum stress tensor operator. In section \ref{sec6} our
validity criterion is applied to the study of flat spacetime as a
solution of semiclassical gravity. The problem of the runaway
solutions and methods to deal them is discussed. Throughout the
paper in order to emphasize the qualitative aspects
we use a simplified notation without
tensorial indices and for a few points we also use qualitative arguments
and order of magnitude estimates. We refer the reader to the
papers \cite{hu03a,hu04a,HuRouVer04a,HuRouVer04b} were the
technical details, as well as many subtleties that cannot be
summarized here, are provided. Our metric and curvature
conventions are those of Ref. \cite{misner73}, and we use
$\hbar=c=1$.

%%%%%%%%%%%%%%%%%%%%%%%%%%%%%%%%%%%%
\section{Semiclassical gravity}
\label{2}
%%%%%%%%%%%%%%%%%%%%%%%%%%%%%%%%%%%%

At present semiclassical gravity cannot be rigorously derived,
but, it can be formally justified in several ways. One of them is
the leading order in the large $N$ expansion
\cite{hartle81}, where $N$ is the number of independent free
quantum fields which interact with gravity only. In this limit,
after path integration one arrives at a theory in which formally
the gravitational field can be treated as a c-number
and the quantum fields are fully quantized.

Semiclassical gravity can be summarized as follows. Let $g$ be the
metric tensor and $\hat\phi$ a scalar field operator. The
semiclassical Einstein equation as the dynamical equation that
describes the back-reaction of quantum matter on the metric
$g$ can be written as
\begin{equation}
G_g=\kappa \langle \hat T^R\rangle_g,
\label{1.1}
\end{equation}
where $\hat T=T[\hat\phi^2]$ is the matter stress tensor in a simplified
notation, which is quadratic in the field operator $\hat \phi$,
and $\kappa=8\pi G $, where $G$ is Newton's constant. This
operator, being the product of distribution valued operators, is
ill defined and needs to be regularized and renormalized, the $R$
in $\hat T^R$ means that the operator has been renormalized. The
angle brackets on the right hand side mean that the expectation
value of the stress tensor operator is computed in some quantum
state, say $|\psi\rangle$, compatible with the geometry described
by the metric $g$. On the left hand side $G_g$ stands for the
Einstein tensor of the metric $g$ together with the cosmological
constant term and other terms quadratic in the curvature which are
generally needed to renormalize the matter stress tensor operator.
The quantum field operator $\hat\phi$ propagates in the background
defined by the metric $g$, it thus satisfies a Klein-Gordon
equation,
\begin{equation}
(\Box_g -m^2)\hat\phi=0,
\label{1.2}
\end{equation}
where $\Box_g$ stands for the D'Alambert operator in the
background of $g$ and $m$ is the mass of the scalar field. A
solution of semiclassical gravity consists of the set $(g,\hat
\phi,|\psi\rangle)$ where $g$ is a solution of Eq.~(\ref{1.1}),
$\hat \phi$ is a solution of Eq.~(\ref{1.2}) and $|\psi\rangle$ is
the quantum state in which the expectation value of the stress
tensor in Eq.~(\ref{1.1}) is computed.

As we recalled in the introduction this theory is in some sense
unique as a theory that describes the interaction of a classical
gravitational field with quantum matter. As an effective theory it
should break down at Planck scales. Also, from the right hand side
of the semiclassical Einstein equation it seems clear that the
theory should also break down when the fluctuations of the quantum
stress tensor are large. This has been emphasized by Ford and
collaborators, and may be illustrated by the example of Ref.
\cite{ford82} as follows.

Let us assume a quantum state formed by an isolated system which
consists of a superposition with equal amplitude of one
configuration with mass $M_1$ and another with mass $M_2$.
Semiclassical theory as described in Eq. (\ref{1.1}) predicts that
the gravitational field of this system is produced by the average
mass $(M_1+M_2)/2$, that is a test particle will move on the
background spacetime produced by such a source. However one would
expect that if we send a succession of test particles to probe the
gravitational field of the above system half of the time they
would react to the field of a mass $M_1$ and the other half to the
field of a mass $M_2$. If the two masses differ substantially the
two predictions are clearly different, note that the fluctuations
in mass of the quantum state is of the order of $(M_1-M_2)^2$.
Although the previous example is suggestive a word of caution
should be said in order not to take it too literary. In fact, if
the previous masses are macroscopic the quantum system decoheres
very quickly \cite{Zur91} and instead of a pure quantum state it
is described by a density matrix which diagonalizes in a certain
pointer basis. Thus for observables associated to this pointer
basis the matrix density description is  equivalent to that
provided by a statistical ensemble. In any case, however, from the
point of view of the test particles the predictions differ from
that of the semiclassical theory.

%%%%%%%%%%%%%%%%%%%%%%%%%%%%%%%%%%%%%%
\section{Stochastic gravity}
\label{3}
%%%%%%%%%%%%%%%%%%%%%%%%%%%%%%%%%%%%%%

The purpose of stochastic (semiclassical) gravity is to be able to
deal with the situation of the previous example when the
predictions of the semiclassical theory may be inaccurate.
Consequently, our first point is to characterize the quantum
fluctuations of the stress tensor.

The physical observable that measures these fluctuations is
$\langle \hat T^2\rangle-\langle \hat T\rangle^2$. To make this
more precise let us introduce the tensor operator $\hat t\equiv\hat T- \langle
\hat T\rangle\hat I$, where $\hat I$ is the identity operator, then we
introduce the {\it{noise kernel}} as the four-index bi-tensor defined as
the expectation value of the anticommutator of the operator $\hat t$:
\begin{equation}
N(x,y)=\frac{1}{2}\langle\{ \hat t(x),\hat t(y)\} \rangle_g.
\label{1.3}
\end{equation}
Thus, the noise kernel is the symmetrized connected part of the
two-point quantum correlation function of the stress tensor
operator with respect to the state of the matter fields. The
subindex $g$ here means that this expectation value in taken in a
background metric $g$. An important property of the symmetric
bi-tensor $N(x,y)$ is that it is finite  because the tensor
operator $\hat t$ is finite since the ultraviolet divergences of
$\hat T$ are cancelled by the substraction of $\langle \hat
T\rangle$. Since the operator $\hat T$ is selfadjoint $N(x,y)$,
which is the expectation value of an anticommutator, is real and
positive semi-definite \cite{hu03a}. Thus, when considering the
inverse kernel $N^{-1}(x,y)$, one must work in the subspace
obtained from the eigenvectors which have strictly positive
eigenvalues when the noise kernel is diagonalized. The last
property allows for the introduction of a classical Gaussian
stochastic tensor $\xi$ defined by
\begin{equation}
\langle\xi(x)\rangle_s=0,\ \ \ \langle\xi(x)\xi(y)\rangle_s=N(x,y).
\label{1.4}
\end{equation}
This stochastic tensor is symmetric and divergenceless,
$\nabla\cdot \xi=0$, as a consequence of the fact that the stress
tensor operator is divergenceless. The subindex $s$ means that the
expectation value is just a classical stochastic average. Note
that we assume that $\xi$ is Gaussian just for simplicity in order
to include the main effect of the quantum fluctuations.

The idea now is simple we want to modify the semiclassical
Einstein equation (\ref{1.1}) by introducing a linear correction
to the metric tensor $g$, such as $g+h$, which accounts
consistently for the fluctuations of the stress tensor. The
simplest equation is,
\begin{equation}
G_{g+h}=\kappa (\langle \hat T^R\rangle_{g+h}+\xi),
\label{1.5}
\end{equation}
where $g$ is assumed to be a solution of equation (\ref{1.1}).
This stochastic equation must be thought of as a linear equation
for the metric perturbation $h$ which will behave, consequently,
as a stochastic field tensor. Note that the tensor $\xi$ is not a
dynamical source, since it has been defined in the background
metric $g$ which is a solution of the semiclassical equation. Note
also that this source is divergenceless with respect to the
metric, and it is thus consistent to write it on the right hand
side of the Einstein equation. This equation is gauge invariant
with respect to diffeomorphisms defined by any field on the
background spacetime \cite{martin99b}. If we take the statistical
average of equation~(\ref{1.5}) it becomes just the semiclassical
equation for the perturbed metric $g+h$ where now the expectation
value of $\hat T$ is taken in the perturbed spacetime.

The stochastic equation (\ref{1.5}) is known as the
Einstein-Langevin equation. To linear order in $h$ we have
\cite{martin99b},
\begin{equation}
\langle \hat T^R\rangle_{g+h}(x)=-2\int H(x,x^\prime)\cdot
h(x^\prime), \label{1.5a}
\end{equation}
where the kernel $H(x,x^\prime)$ has three terms, one of them is
proportional to the imaginary part of the expectation value of the
time ordered two-point stress tensor, $\mathrm{Im}\langle T(\hat
T(x)\hat T(x^\prime))\rangle$, the second term is proportional to
the expectation value of the stress tensor commutator, $\langle
[\hat T(x),\hat T(x^\prime)]\rangle$, and the third is
proportional to the functional derivative of $\langle \hat
T\rangle $ with respect to the metric (excluding the implicit
dependence on the metric of the field $\hat\phi$).
Of course, this kernel is
also the main ingredient of the linearized semiclassical Einstein
equation around a given background metric $g$. The other key
ingredient in the Einstein-Langevin equation is the noise kernel
$N(x,y)$ which defines the stochastic inhomogeneous source of the
equation. This kernel should be thought of as a distribution
function, the limit of coincidence points has meaning only in the
sense of distributions. Explicit expressions of this kernel in
terms of the two point Wightman functions are given in Ref.
\cite{martin99b} on a general background. Detailed expressions for
this kernel in the Minkowski background are given in Ref.
\cite{martin00}, and expression based on point-splitting methods
have also been given in Refs. \cite{RouVer00,phillips00} in other
backgrounds.

The Einstein-Langevin equation has been previously derived making
use of a formal analogy with open quantum systems and employing
the influence functional formalism \cite{feynman63,feynman65}. The
basis for this approach is a functional formalism known as closed
time path, first introduced by Schwinger
\cite{schwinger61,keldysh65,chou85}, which is an effective action
method suitable to derive dynamical equations for expectation
values of quantum operators; rather than transition elements as in
the standard effective action method. The closed time path
formalism was later applied to the problem of back-reaction of
quantum fields on the spacetime metric
\cite{jordan86,calzetta87,campos94}, in order to derive
semiclassical Einstein equations. The formalism was then applied
along the lines of the influence functional formalism to derive
Einstein-Langevin equations in several contexts
\cite{calzetta94,hu95a,hu95b,campos96,calzetta97c,martin99b,martin99c}.
In Ref.~\cite{martin99a} the Einstein-Langevin equation was
derived by an axiomatic approach by arguing that it is the only
consistent generalization of the semiclassical Einstein equation
which takes into account the back-reaction of the
matter stress tensor fluctuations to lowest order.
We have summarized the axiomatic approach in this section.

The solution of the Einstein-Langevin equation (\ref{1.5}), taking
into account Eq. (\ref{1.5a}), may be expressed as,
\begin{equation}
h(x)=h^0(x)+\kappa\int G_{R}(x,x^\prime)\cdot \xi(x^\prime),
\label{1.5b}
\end{equation}
where $h^0$ is a solution of the homogeneous part of equation
(\ref{1.5})  which contains all the information on the initial
conditions, and $G_{R}(x,x^\prime)$ is the retarded propagator
with vanishing initial conditions associated with the equation
(\ref{1.5}). The two-point correlation function for the metric
perturbation which is the physically most relevant observable can
then be written as:
\begin{eqnarray}
&&\langle h(x)h(y)\rangle_s =\langle h^0(x)h^0(y)\rangle_s +
\nonumber\\
&&\quad\quad\quad\quad\quad\kappa^2 \int G_R(x,x^\prime)\cdot
N(x^\prime,y^\prime)\cdot G_R(y,y^\prime), \label{1.5c}
\end{eqnarray}
where the first average, $\langle h^0(x)h^0(y)\rangle_s$, is
taken with respect to the initial conditions.

It turns out that going to leading order in $1/N$, in the large
$N$ expansion, one can show that the stochastic correlation
functions for the metric perturbations obtained from the
Einstein-Langevin equation coincide with the symmetrized two-point
quantum correlation functions of the metric perturbations. The
details of the derivation will be given in Ref.~\cite{roura03b}
and are summarized in Ref.~\cite{HuRouVer04a} for the particular
case of a Minkowski background, to which we will restrict in
section \ref{sec6}. In this case $\kappa$ in Eq. (\ref{1.5b}) has to be
replaced by the rescaled gravitational coupling constant
$\bar{\kappa} = N \kappa$ and the noise kernel for a single field
$N(x,y)$ must be replaced by $(1/N)N(x,y)$. Thus, we have that the
symmetrized two-point quantum correlation function for the metric
perturbation is
\begin{equation}
\frac{1}{2}\langle\{\hat h(x),\hat h(y)\}\rangle=\langle h(x)h(y)\rangle_s.
\label{1.5d}
\end{equation}
where the Lorentz gauge condition $\nabla\cdot (h - (1/2) \eta
\mathrm{Tr}h) = 0$ ($\eta$ is the Minkowski metric) as well as
some initial condition to fix completely the remaining gauge
freedom of the initial state should be implicitly understood.

It should be emphasized that there are two different contributions to
the symmetrized quantum correlation function, which are clearly
distinguished in Eq.~(\ref{1.5c}). The first contribution is
related to the quantum fluctuations of the initial state of the
metric perturbations and corresponds to the so called
\emph{intrinsic} fluctuations; here the stochastic average must be taken
with respect to the Wigner distribution function
that describes the initial quantum state. The second contribution is
proportional to the noise kernel, it accounts for the fluctuations of
the stress tensor of the matter fields and corresponds to the so
called \emph{induced} fluctuations. These two contributions to the
two-point correlation functions  is also seen in the description
of some quantum Brownian motion models which are typically used as
paradigms of open quantum systems
\cite{calzetta03a,CalRouVer01,CalRouVer02}. Both, the
intrinsic and induced fluctuations, play a role in our stability
criterion for the solutions of semiclassical gravity.

The full two-point quantum correlation function
for the metric $\langle\hat h(x)\hat h(y)\rangle$ can, in fact, be obtained
{}from the Einstein-Langevin equation. Since this correlation can be
given in terms of the antisymmetrized and the symmetrized quantum
correlation function we only need the commutator that to leading
order in $1/N$ is independent of the initial state of the metric
perturbation and is given by
\begin{equation}
\frac{1}{2}\langle[\hat h(x),\hat h(y)]\rangle=i\kappa
[G_R(y,x)-G_R(x,y)]. \label{1.5e}
\end{equation}
Note that the information on the retarded propagator is already in
the linearized semiclassical Einstein equation. That is, Eq.
(\ref{1.5}) without the stochastic source.

%%%%%%%%%%%%%%%%%%%%%%%%%%%%%
\subsection{A toy model}
%%%%%%%%%%%%%%%%%%%%%%%%%%%%%

To justify Eq. (\ref{1.5d}) which plays an essential role in our
criteria for the validity of semiclassical gravity it is useful to
introduce a simple toy model for gravity which minimizes the
technical complications. The model is also useful to clarify the
role of the noise kernel and illustrate the relationship between
the semiclassical, stochastic and quantum descriptions. Let us
assume that the gravitational equations are described by a
massless scalar field $h$ whose source is another massless scalar
field $\phi$ which satisfies the Klein-Gordon equation in flat
spacetime $\Box \phi=0$. The field stress tensor is quadratic in
the field, and independent of $h$. The classical gravitational
field equations will be given by
\begin{equation} \Box h=\kappa T,
\label{2.12a}
\end{equation}
where $T$ is now the (scalar) trace of the stress tensor. Note that
this is not a self-consistent theory since $\phi$ does not react
to the gravitational field $h$. This model obviously differs from the
standard linearized theory of gravity discussed previously, where
$T$ is also linear in $h$, but it captures some of its key
features.

In the Heisenberg representation the quantum scalar field $\hat h$ satisfies
\begin{equation}
\Box \hat h=\kappa\hat T.
\label{2.12}
\end{equation}
Since $\hat T$ is quadratic in the field operator $\hat\phi $ some
regularization procedure has to be assumed in order for Eq.
(\ref{2.12}) to make sense. Since we work in flat spacetime we may
simply use a normal ordering prescription to regularize the
operator $\hat T$. The solutions of this equation, i.e. the field
operator at the point $x$, which we call $\hat h_x$ in this
subsection to avoid confusion with
the more standard notation, $\hat h(x)$,
used in the rest of the paper, may be written in terms of the
retarded propagator $G_{xx^\prime}$ of the D'Alambertian as,
\begin{equation} \hat h_x=\hat
h^0_x+\kappa\int G_{xx^\prime}\hat T_{x^\prime}, \label{2.13}
\end{equation}
where $\hat h^0_x$ is the free field which carries information on
the initial conditions and the state of the field. {}From this
solution we may compute, for instance, the symmetric two-point
quantum correlation function (the anticommutator)
\begin{equation}
\langle \{\hat h_x,\hat h_y\}\rangle = \langle \{\hat h^0_x,\hat
h^0_y\}\rangle + \kappa^2 \int G_{xx^\prime}G_{yy^\prime}
\langle\{\hat T_{x^\prime},\hat T_{y^\prime}\}\rangle,
\label{2.14}
\end{equation}
where the expectation value is taken with respect to the quantum
state in which both fields $\phi$ and $h$ are quantized.
We have assumed $\langle \hat h^0\rangle=0$ for the free
field.

We can now consider the semiclassical theory for this problem. If
we assume that $h$ is classical and the matter field is quantum
the semiclassical limit may just be described by substituting
into the classical equation (\ref{2.12a}) the stress trace by the
expectation value of the scalar stress operator $\langle \hat
T\rangle$, in some quantum state of the field $\hat \phi$. We may simply
renormalize the expectation value of $\hat T$ using normal ordering, then for
the vacuum state of the field $\hat\phi$, we would simply have
$\langle\hat T\rangle_0=0 $. The semiclassical theory thus reduces
to
\begin{equation}
\Box  h=\kappa \langle \hat T\rangle.
\label{2.15a}
\end{equation}
The two point function $h_xh_y$ that one may derive from this
equation depends on the two point function $\langle \hat
T_x\rangle \langle \hat T_y\rangle $ and clearly cannot reproduce
the quantum result of Eq.~(\ref{2.14}) which depends on the
expectation value of two-point operator $\langle\{\hat T_x,\hat
T_y\}\rangle$. That is, the semiclassical theory entirely misses
the fluctuations of the scalar stress operator $\hat T$.

To extend this semiclassical theory in order to account for such
fluctuations, we introduce the noise kernel as we did in the
previous section. Thus, we define
\begin{equation}
N_{xy}= \frac{1}{2}\langle\{\hat
t_x,\hat t_y\}\rangle
\label{2.14a}
\end{equation}
where $\hat t\equiv\hat T-\langle\hat T\rangle$, and we have used again
the sub-index notation to avoid confusion with the noise kernel of
the previous section. The bi-scalar $N_{xy}$ is real and
positive-semidefinite, as a consequence of $\hat t$ being
self-adjoint \cite{hu03a}. Consequently we can introduce a
Gaussian stochastic field as:
\begin{equation}
\langle\xi\rangle_s=0,\quad \langle\xi_x\xi_y\rangle_s=N_{xy}.
\label{2.14b}
\end{equation}
where the subscript $s$ means a statistical average.

The extension of the semiclassical equation may be simply
performed by adding to the right-hand side of the semiclassical
equation (\ref{2.15a}) the stochastic source $\xi$, which accounts
for the fluctuations of $\hat T$ as follows,
\begin{equation}
\Box  h=\kappa\left( \langle \hat T\rangle+\xi\right).
\label{2.15}
\end{equation}
This equation is in the form of a Langevin equation: the field $h$
is classical but stochastic and the observables we may obtain from
it are correlation functions for $h$. In fact, the solution of
this equation may be written in terms of the retarded propagator
as, \begin{equation} h_x=h^0_x+\kappa\int
 G_{xx^\prime}\left(\langle\hat T_{x^\prime}\rangle
+\xi_{x^\prime}\right) , \label{2.16}
\end{equation}
{}from where the two point correlation function for the classical field
$h$, after using the definition of $\xi$ and that $\langle
h^0\rangle_s=0$, is given by
\begin{equation}
\langle  h_x h_y\rangle_s = \langle  h^0_x h^0_y\rangle_s
+\frac{\kappa^2}{2}\int G_{xx^\prime}G_{yy^\prime}\langle\{\hat
T_{x^\prime},\hat T_{y^\prime}\}\rangle. \label{2.17}
\end{equation}
Note that in writing $\left<\dots\right>_s$ here we are assuming a
double stochastic average, one is related to the stochastic
process $\xi$ and the other is related to the free field $h^0$
which is assumed also to be stochastic with an initial
distribution function to be specified.

Comparing Eqs.~(\ref{2.14}) and (\ref{2.17}) we see that the
respective second term on the right-hand side are identical
(except for a factor of $2$ due to the symmetrization) provided
the expectation values are computed in the same quantum state for
the field $\hat \phi$. The fact that the field $h$ is also quantized
in (\ref{2.14}) does not change the previous statement; recall that
$T$ does not depend on $h$. The nature
of the first term on the right-hand sides of equations
(\ref{2.14}) and (\ref{2.17}) is different: in the first case it
is the two-point quantum expectation value of the free quantum
field $\hat h^0$ whereas in the second case it is the stochastic
average of the two point classical homogeneous field $h^0$, which
depends on the initial conditions. Now we can still make these
terms equal to each other (with the factor of $2$) if we assume
for the homogeneous field $h^0$ a Gaussian distribution of initial
conditions such that
\begin{equation}
\langle  h^0_x
h^0_y\rangle_s=  \frac{1}{2}\langle\{\hat h^0_x,\hat
h^0_y\}\rangle.
\label{2.17a}
\end{equation}
This Gaussian stochastic field $h^0$ can always be defined due to
the semi-positivity of the anti-commutator. Thus, under this
assumption on the initial conditions for the field $h$ the two
point correlation function of Eq.~(\ref{2.17}) equals the quantum
expectation value of Eq.~(\ref{2.14}) exactly. Thus, we have
\begin{equation}
\frac{1}{2}\langle \{\hat h_x,\hat h_y\}\rangle=\langle h_x
h_y\rangle_s,
\label{2.17b}
\end{equation}
which may be compared to Eq. (\ref{1.5d}). 
Comparing with  the linearized theory of gravity described in
the previous section we see that $\langle T\rangle$ depends also
on $h$, both explicitly and also implicitly through the coupling
of $\phi$ with $h$. The retarded propagator here $G_{xx^\prime}$
is then replaced by the propagator $G_R(x,x^\prime)$ of the
previous section and the functions $h^0$, which are here the free
metric perturbations are replaced by the homogeneous solutions of
the previous section.

%%%%%%%%%%%%%%%%%%%%%%%%%%%%%%%%%%%%%%%%%%%
\section{Stability of flat spacetime}
\label{sec6}
%%%%%%%%%%%%%%%%%%%%%%%%%%%%%%%%%%%%%%%%%%%

Let us now apply our validity criterion to flat spacetime. One
particularly simple and interesting solution of semiclassical
gravity is the Minkowski metric. In fact, when the quantum fields
are in the Minkowski vacuum state one may take the renormalized
expectation value of the stress tensor $\langle T^R\rangle=0$
(this is equivalent to assuming that the cosmological constant is
zero) and the Minkowski metric $\eta$ is a solution of the
semiclassical Einstein equation (\ref{1.1}). Thus, we can look for
the stability of flat spacetime against quantum matter fields.
According to the criteria we have established we have to look for
the behavior of the two-point quantum correlations for the metric
perturbations $h$ over the Minkowski background which are given by
Eqs.~(\ref{1.5c}) and (\ref{1.5d}). As we have emphasized several
times these fluctuations separate in two parts: the first term on
the right hand side of Eq.~(\ref{1.5c}) corresponds to the
\textit{intrinsic} fluctuations, and the second term corresponds
to the \textit{induced} fluctuations.

\subsection{Intrinsic fluctuations}

Let us first consider the intrinsic fluctuations,
\begin{equation}
\langle h^0(x) h^0(y)\rangle_s, \label{6.1}
\end{equation}
where $h^0$ are the homogeneous solutions of the
Einstein-Langevin equation (\ref{1.5}), or equivalently the
linearly perturbed semiclassical equation, and the statistical
average is taken with respect to the Wigner distribution that
describes the initial quantum state of
the metric perturbations. Since these solutions are
described by the linearized semiclassical equation around flat
spacetime we can make use of the results derived in
Refs.~\cite{horowitz80,flanagan96,anderson03}. The solutions for
the case of a massless scalar field were first discussed in
Ref.~\cite{horowitz80} and an exhaustive description can be found
in Appendix~A of Ref.~\cite{flanagan96}. Decomposing the metric
perturbation into scalar, vectorial and tensorial parts and
computing the linearized Einstein tensor, one gets a vanishing
result for the vectorial part of the metric perturbation; the
scalar and tensorial components of the metric perturbation give
rise, respectively, to the scalar and tensorial components of the
linearized Einstein tensor.
The vectorial part is found to vanish whereas the scalar
and tensorial contributions for a massless and conformally coupled
scalar field (see Ref.~\cite{flanagan96} for the massless case
with arbitrary coupling and Refs.~\cite{martin00,anderson03} for
the general massive case) satisfy the following equations:
\begin{equation}
\left( 1 + 12 \kappa \bar{\beta} p^2 \right) \tilde{G}^{
\mathrm{(S)}}(p) = 0 \label{scalar},
\end{equation}
\begin{equation}
\lim\limits_{\;\epsilon \rightarrow 0^+} \left( 1 + \frac{\kappa
p^2}{960 \pi^2}  \ln  \frac{p^2 }{\mu^2} \right) \tilde{G}^{
\mathrm{(T)} }(p) = 0 \label{tensorial},
\end{equation}
where in the last equation the prescription that the time
component of $p$ has a small imaginary part, $p^0+i \epsilon$, is
taken. Here $\tilde G(p)$ stands for the Fourier transform of the
linearized Einstein tensor, the upper indices $S$ and $T$ stand
for scalar and tensorial respectively, $\bar\beta$ is a
dimensionless renormalized parameter that multiplies some of 
the quadratic terms in the curvature in the effective action for the
gravitational field, and $\mu$ is a renormalization mass scale. 
See Ref.~\cite{HuRouVer04a} for a more complete description.

For the scalar component when $\bar{\beta} = 0$ the only solution
is $\tilde{G}^{ \mathrm{(S)} }(p) = 0$. When $\bar{\beta}
> 0$ the solutions for the scalar component exhibit an oscillatory
behavior in spacetime coordinates which corresponds to a massive
scalar field with $m^2 = (12 \kappa |\bar{\beta}|)^{-1}$; for
$\bar{\beta} < 0$ the solutions correspond to a tachyonic field
with $m^2 = - (12 \kappa |\bar{\beta}|)^{-1}$: in spacetime
coordinates they exhibit an exponential behavior in time, growing
or decreasing, for wavelengths larger than $4 \pi (3 \kappa
|\bar{\beta}|)^{1/2}$, and an oscillatory behavior for wavelengths
smaller than $4 \pi (3 \kappa |\bar{\beta}|)^{1/2}$. On the other
hand, the solution $\tilde{G}^{ \mathrm{(S)} }(p) = 0$ is
completely trivial since any scalar metric perturbation
$\tilde{h}(p)$ giving rise to a vanishing linearized Einstein
tensor can be eliminated by a gauge transformation as explained in
Ref. \cite{HuRouVer04a}.

For the tensorial component, when $\mu \le \mu_\mathrm{crit} =
l_p^{-1} (120\pi)^{1/2} e^{\gamma}$, where $l_p$ is the Planck length
($l_p^2\equiv \kappa/8\pi$) the first factor in
Eq.~(\ref{tensorial}) vanishes for four complex values of $p^0$ of
the form $\pm \omega$ and $\pm \omega^*$, where $\omega$ is some
complex value. We will consider here the case in which $\mu <
\mu_\mathrm{crit}$; a detailed description of the situation for
$\mu \ge \mu_\mathrm{crit}$ can be found in Appendix~A of
Ref.~\cite{flanagan96}. The two zeros on the upper half of the
complex plane correspond to solutions in spacetime coordinates
exponentially growing in time, whereas the two on the lower half
correspond to solutions exponentially decreasing in time. Strictly
speaking, these solutions only exist in spacetime coordinates,
since their Fourier transform is not well defined. They are
commonly referred to as runaway solutions and for $\mu \sim
l_p^{-1}$ they grow exponentially in time scales comparable to the
Planck time.

In order to deal with those unstable solutions, one possibility is
to employ the \textit{order reduction} prescription
\cite{parker93}, which we will briefly summarize in the last
subsection. With such a prescription we are left only with the
solutions which satisfy $\tilde{G}(p)=0$. The solutions for
$\tilde{h}(p)$ simply correspond to free linear gravitational
waves propagating in Minkowski spacetime expressed in the
transverse and traceless (TT) gauge. When substituting back into
Eq.~(\ref{6.1}) and averaging over the initial conditions we
simply get the symmetrized quantum correlation function for free
gravitons in the TT gauge for the state given by the
Wigner distribution. As far as the intrinsic fluctuations are
concerned, it seems that the order reduction prescription is too
drastic, at least in the case of Minkowski spacetime, since no
effects due to the interaction with the quantum matter fields are
left.

A second possibility, proposed by Hawking \emph{et al.}
\cite{hawking01,hawking02}, is to impose boundary conditions which
discard the runaway solutions that grow unbounded in time and
correspond to a special prescription for the integration contour
when Fourier transforming back to spacetime coordinates. Following
that procedure we get, for example, that for a massless
conformally coupled scalar field with  $\bar{\beta} > 0$ the
intrinsic contribution to the symmetrized quantum correlation
function coincides with that of free gravitons plus an extra
contribution for the scalar part of the metric perturbations which
renders Minkowski spacetime stable but plays a crucial role in
providing a graceful exit for inflationary models driven by the
vacuum polarization of a large number of conformal fields. Such a
massive scalar field would not be in conflict with present
observations because, for the range of parameters
considered, the mass would be far too large to have observational
consequences \cite{hawking01}.

\subsection{Induced fluctuations}

The induced fluctuations are described by the second term in
Eq.~(\ref{1.5c}). They are induced for the noise kernel that describes
the stress tensor fluctuations of the matter fields,
\begin{equation}
\frac{\bar\kappa^2}{N}\int G_R(x,x^\prime)\cdot N(x^\prime,y^\prime)\cdot
G_R(y,y^\prime) , \label{6.2}
\end{equation}
where we write the expression in the large $N$ limit. The
contribution corresponding to the induced quantum fluctuations is
equivalent to the stochastic correlation function obtained by
considering just the inhomogeneous part of the solution to the
Einstein-Langevin equation: the second term on the right-hand side
of Eq.~(\ref{1.5c}). Taking all that into account, it is clear
that we can make use of the results for the metric correlations
obtained in Ref.~\cite{martin00} by solving the Einstein-Langevin
equation. In fact, one should simply take $N=1$ to transform our
expressions to those of Ref.~\cite{martin00} and, similarly,
multiply the noise kernel in the expressions of that reference by
$N$ so that they can be used here, which follows
{}from the fact that we have $N$ independent matter fields.

The same kind of exponential instabilities in the runaway
solutions of the homogeneous part of the Einstein-Langevin
equation also arise when computing the retarded propagator
$G_\mathrm{R}$. In order to deal with those instabilities, similar
to the case of the intrinsic fluctuations, one possibility is to
make use of the order reduction prescription. The
Einstein-Langevin equation becomes then $\tilde G (p)=
\bar\kappa\tilde \xi (p)$. The second possibility, following the
proposal of Hawking \emph{et al.}, is to impose boundary
conditions which discard the exponentially growing solutions and
translate into a special choice of the integration contour when
Fourier transforming back to spacetime coordinates the expression
for the propagator. In fact, it turns out that the propagator
which results from adopting that prescription coincides with the
propagator that was employed in Ref.~\cite{martin00}. Note,
however, that this propagator is no longer a strictly retarded
propagator since it exhibits causality violations at Planck
scales. A more detailed discussion on all these points can be
found in Appendix E of Ref.~\cite{HuRouVer04a}.

Following Ref.~\cite{martin00}, the Einstein-Langevin equation can
be entirely written in terms of the linearized Einstein tensor.
The equation involves second derivatives of that tensor, and  in
terms of its Fourier components $\tilde{G}(p)$ takes the form
\begin{equation}
(1+ F(p))\cdot \tilde{G}(p) = \bar{\kappa} \tilde{\xi} (p) ,
\label{6.3}
\end{equation}
where $F$ is a four-index tensor which depends on $p^2\ln p^2$
when the field is massless and conformally coupled. This reflects
the fact that we have second derivatives of the Einstein tensor
and the nonlocality of the Einstein-Langevin equation (or also of
the perturbed semiclassical equation). {}From equation (\ref{6.3})
one may obtain the correlation functions for $\tilde{G}(p)$,
$\langle \tilde G (p)\tilde G (q)\rangle_s$, which are invariant
under gauge transformations of the metric perturbations. Writing
the linearized Einstein tensor in terms of the metric
perturbation, which takes a particularly simple form in the
Lorentz gauge, one may derive the correlation functions for
$\tilde{h}(p)$: $\langle \tilde h (p)\tilde h (q)\rangle_s$.
Finally, the correlation functions in spacetime coordinates can be
easily obtained by Fourier transforming these correlations. For
massless and conformally coupled matter fields explicit results
are given in Ref.~\cite{martin00}, they have the general
expression:
\begin{equation}
\langle h(x)h(y)\rangle_s=\frac{\bar\kappa^2}{720\pi N}\int
\frac{e^{ip\cdot
(x-y)}P\theta(-p^2)}{|1+(\bar\kappa/960\pi^2)p^2\ln(p^2/\mu^2)|^2}
 \label{6.4}
\end{equation}
where $P$ is a four-index projection tensor. This correlation
function for the metric perturbations is in agreement with the
real part of the propagator obtained by Tomboulis in
Ref.~\cite{tomboulis77} using a large $N$ expansion.

To estimate this integral let us consider spacelike separated
points $x-y=(0,\mathbf{r})$ and introduce the Planck length
$l_p$. It is not difficult to see \cite{hu04a}, that for space
separations $|\mathbf{r}|\gg l_p$ we have
\begin{equation}
\langle h(x)h(y)\rangle_s\sim \frac{l_p^4}{|\mathbf{r}|^4},
 \label{6.5}
\end{equation}
and for $|\mathbf{r}|\sim N l_p$ we have
\begin{equation}
\langle h(x)h(y)\rangle_s\sim
e^{-|\mathbf{r}|/l_p}\frac{l_p}{|\mathbf{r}|}.
 \label{6.6}
\end{equation}
Since these fluctuations are induced by the matter stress fluctuations we
infer that the effect of the matter fields is to suppress metric
fluctuations at small scales. On the other hand, at large scales
the induced metric fluctuations are small compared to the free
graviton propagator which goes like $l_p^2/|\mathbf{r}|^2$.

We thus conclude that, once the instabilities giving rise to the
unphysical runaway solutions have been discarded, the fluctuations
of the metric perturbations around the Minkowski spacetime induced
by the interaction with quantum scalar fields are indeed stable.
Instabilities would lead to a divergent result when Fourier
transforming back to spacetime coordinates. Note that when the
order reduction prescription is used the $p^2\ln p^2$ terms are
absent in the corresponding Eq. (\ref{6.4}). Thus, in contrast to
the intrinsic fluctuations, there is still a nontrivial
contribution to the induced fluctuations due to the quantum matter
fields in this case.

\subsection{Order reduction prescription and large $N$}

Runaway solutions are a typical feature of equations describing
back-reaction effects, such is in classical electrodynamics, and
are due to higher than two time derivatives in the dynamical
equations. In a very schematic way the semiclassical Einstein
equations have the form
\begin{equation}
G_h+l_p^2\ddot G_h=0,
\label{6.7}
\end{equation}
where $G_h$ stands for the linearized Einstein tensor
over the Minkowski background, say,
and we have simplified the equation as much as
possible. The second term of the equation is due to the vacuum
polarization of matter fields and contains four time derivatives
of the metric perturbation. Some specific examples of such an equation
are, in momentum space, Eqs.~(\ref{scalar}) and (\ref{tensorial}).
The order reduction
procedure is based on treating perturbatively the terms involving
higher order derivatives, differentiating the equation under
consideration and substituting back the higher derivative terms in
the original equation keeping only terms up to the required order
in the perturbative parameter. In the case of the semiclassical
Einstein equation, the perturbative parameter is $l_p^2$. If we
differentiate twice Eq.~(\ref{6.7}) with respect to time it is
clear that the second order derivatives of the Einstein tensor are
of order $l_p^2$. Substituting back into the original equation, we
get the following equation up to order $l_p^4$: $ G_h=0+ O(l_p^4). $
Now, there are certainly no runaway solutions but also no effect
due to the vacuum polarization of matter fields. Note that the
result is not so trivial when there is an inhomogeneous term on
the right hand side of Eq.~(\ref{6.7}), this is what happens with the
induced fluctuations predicted by the Einstein-Langevin equation.

Semiclassical gravity is expected to provide reliable results as
long as the characteristic length scales under consideration, say
$L$, satisfy that $L\gg l_p$ \cite{flanagan96}. This can be
qualitatively argued by estimating the magnitude of the different
contributions to the effective action for the gravitational field,
considering the relevant Feynman diagrams and using dimensional
arguments. Let us write the effective gravitational action, again
in a very schematic way, as
\begin{equation}
S_{\mathrm{eff}}=\int \sqrt{-g}\left( \frac{1}{l_p^2}R
+\alpha R^2+l_p^2 R^3+\dots \right),
\label{6.8}
\end{equation}
where $R$ is the Ricci scalar. The first term is the usual
classical Einstein-Hilbert term, the second stands for terms
quadratic in the curvature (square of Ricci and Weyl tensors) this
terms appear as radiative corrections due to vacuum polarization
of matter fields, here $\alpha$ is an dimensionless parameter
presumably of order 1, the $R^3$ terms are higher order
corrections which appear for instance when one considers internal
graviton propagators inside matter loops. Let us assume that
$R\sim L^{-2}$ then the different terms in the action are of the
order of $R^2\sim L^{-4}$ and $l_p^2R^3\sim l_p^2L^{-6}$.
Consequently when $L\gg l_p^2$, the term due to matter loops is a
small correction to the Einstein-Hilbert term $(1/l_p^2)R\gg R^2$,
and this term can be treated as a perturbation. The justification
of the order reduction prescription is actually based on this
fact. Therefore, significant effects from the vacuum polarization
of the matter fields are only expected when their small
corrections accumulate in time, as would be the case, for
instance, for an evaporating macroscopic black hole all the way
before reaching Planckian scales.

However if we have a large number $N$ of matter fields the
estimates for the different terms change in a remarkable way. This
is interesting because the large $N$ expansion seems the best
justification for semiclassical gravity. In fact, now the vacuum
polarization terms involving loops of matter are of order
$NR^2\sim NL^{-4}$. For this reason the contribution of the
graviton loops, which is of order $R^2$, can be neglected in front
of the matter loops; this justifies the semiclassical
limit. Similarly higher order corrections are of order
$Nl_p^2R^3\sim Nl_p^2L^{-6}$. Now there is a regime, when $L\sim
\sqrt{N}l_p$, where the Einstein-Hilbert term is comparable to the
vacuum polarization of matter fields, $(1/l_p^2)R\sim NR^2$, and
yet the higher correction terms can be neglected because we still
have $L\gg l_p$, provided $N\gg 1$. This is the kind of situation
considered in trace anomaly driven inflationary models
\cite{hawking01}, such as that originally proposed by Starobinsky
\cite{starobinsky80}, where the exponential inflation is driven by
a large number of massless conformal fields. The order reduction
prescription would completely discard the effect from the vacuum
polarization of the matter fields even though it is comparable to
the Einstein-Hilbert term. In contrast, the procedure proposed by
Hawking \emph{et al.} keeps the contribution from the matter
fields. Note that here the actual physical Planck length $l_p$ is
considered, not the rescaled one, $\bar{l}_p^2 = \bar\kappa/8\pi$,
which is related
to $l_p$ by $l_p^2 = \kappa/8\pi= \bar{l}_p^2/ N$.

%%%%%%%%%%%%%%%%%%%%%
\section{Conclusions}
%%%%%%%%%%%%%%%%%%%%%%

An analysis of the stability of any solution of semiclassical
gravity with respect to small quantum corrections should consider
not only the evolution of the expectation value of the metric
perturbations around that solution, but also their fluctuations,
encoded in the quantum correlation functions. Making use of the
equivalence (to leading order in $1/N$, where $N$ is the number of
matter fields) between the stochastic correlation functions
obtained in stochastic semiclassical gravity and the quantum
correlation functions for metric perturbations around a solution
of semiclassical gravity, the symmetrized two-point quantum
correlation function for the metric perturbations can be
decomposed into two different parts: the intrinsic fluctuations
due to the fluctuations of the initial state of the metric
perturbations itself, and the fluctuations induced by their
interaction with the matter fields. If one considers the
linearized perturbations of the semiclassical Einstein equation,
only information on the intrinsic fluctuations can be retrieved.
On the other hand, the information on the induced fluctuations
naturally follows from the solutions of the Einstein-Langevin
equation.

As a specific example, we have analyzed the symmetrized two-point
quantum correlation function for the metric perturbations around
the Minkowski spacetime interacting with $N$ scalar fields
initially in the Minkowski vacuum state. Once the ultraviolet
instabilities which are ubiquitous in semiclassical gravity and
are commonly regarded as unphysical, have been properly dealt with
by using the order reduction prescription or the procedure
proposed by Hawking \emph{et al.} \cite{hawking01,hawking02}, both
the intrinsic and the induced contributions to the quantum
correlation function for the metric perturbations are found to be
stable \cite{HuRouVer04a}. Thus, we conclude that Minkowski
spacetime is a valid solution of semiclassical gravity.

%%%%%%%%%%%%%%%%%%%%%%%%%
\section{Acknowledgments}
%%%%%%%%%%%%%%%%%%%%%%%%%

I am very grateful to Hans-Thomas Elze and co-organizers for
giving me the opportunity to participate at the Second International Workshop
DICE2004 on {\it
{}From Decoherence and Emergent Classicality to Emergent Quantum Mechanics},
and for their kind and generous hospitality. The research reported
here is mainly based in work made in collaboration with Bei-Lok Hu
and Albert Roura I am grateful to them  for their invaluable
contribution to this topic and for having enjoyed many fruitful discussions.
I thank Daniel Arteaga, Lajos Diosi, Jonathan Halliwell,
Jim Hartle, and Renaud Parentani
for discussions on this topic.
I also thank Albert Roura for a critical reading of the
manuscript and many fruitful discussions. This work has been partially
supported by the MEC Research Projects No. FPA2001-3598 and FPA2004-04582.


\end{document}